\documentclass{revtex4}

\usepackage{amsmath}
\usepackage{ulem}

\begin{document}

\normalem

\title{Lawson Method for Obtaining Wave Functions and $g$ Factors of Ar Isotopes}
\author{L. Zamick}
\affiliation{Department of Physics and Astronomy, Rutgers University
Piscataway, New Jersey 08854}
\author{S. Yeager}
\affiliation{Department of Physics and Astronomy, Rutgers University
Piscataway, New Jersey 08854}
\author{Y. Y. Sharon}
\affiliation{Department of Physics and Astronomy, Rutgers University
Piscataway, New Jersey 08854}
\author{S. J. Q. Robinson}
\affiliation{Department of Physics, Millsaps College
Jackson, Mississippi 39210}

\date{August 13, 2008}

\begin{abstract}
Lawson has shown that one can obtain sensible wave functions even in the weak deformation limit of the Nilsson model as long as one projects out states of good total angular momentum. We apply this method to obtain wave functions and magnetic $g$ factors of excited states of select even-even Ar isotopes.
\end{abstract}
\maketitle

\section{Introduction}

Having participated in large space shell model calculations of $g$ factors of excited states of $^{36, 38, 40}$Ar \cite{stephan05}, \cite{speidel06}, and \cite{speidel08}, we here try to gain insight into the results by using simple wave functions. Furthermore, we extend the calculations to heavier Ar isotopes $^{44}$Ar and $^{46}$Ar.

We will use simple wave functions in which the two proton holes in an Ar isotope are in the $d_{3/2}$ shell and for nuclei like $^{40}$Ar and beyond the valence neutrons are in the $f_{7/2}$ shell. Normally one diagonalizes the matrix hamiltonian with some residual interaction. But the Lawson technique is different \cite{lawson61, lawson80}.

One first forms and intrinsic state by putting the neutrons and protons in single particle Nilsson orbits with the single particle wave functions taken at the weak deformation limit. One then expands the intrinsic state in terms of states of definite total angular momentum. This will be best illustrated by specific examples in the next section.

\section{Parameters}

The Schmidt magnetic moment of a $d_{3/2}$ proton is $0.126 \mu _N$ and that of a $f_{7/2}$ neutron is $-1.913 \mu _N$ (same as a free neutron). However, the measured values are respectively $0.39187 \mu _N$ and $-1.59478 \mu _N$. Hence, we have two sets of $g$ factors.

\begin{center}
Bare

\begin{equation*}
\begin{matrix}
g_{\pi } = 0.0841 & g_{\nu } = -0.5466 \\
& \\
\dfrac{g_{\pi }+g_{\nu }}{2} = -0.2313 & \quad \dfrac{g_{\pi }-g_{\nu }}{2} = 0.3153
\end{matrix}
\end{equation*}

Empirical
\begin{equation*}
\begin{matrix}
g_{\pi } = 0.2610 & g_{\nu } = -0.4557 \\
& \\
\dfrac{g_{\pi }+g_{\nu }}{2} = -0.0973 & \quad \dfrac{g_{\pi }-g_{\nu }}{2} = 0.3583
\end{matrix}
\end{equation*}
\end{center}

\section{The Wave Functions}

\subsection{$^{38}$Ar}

In the the single $j$ shell model the $2_1^+$ states of $^{38}$Ar has the configuration $(d_{3/2})_{\pi }^2J = 2$. The $g$ factor of this state in this model is equal to that of the single hole nucleus $^{39}$K $(J=3/2+)$, namely 0.26098. The measured value for $^{38}$K is +0.24(12), as shown in the work of Speidel et al. \cite{speidel06}. This is very close to the single particle estimate with empirical single particle $g$ factors.

\subsection{$^{40}$Ar Prolate}

Here the configuration is $d_{3/2, \pi }^2 \quad f_{7/2, \nu }^2$. First assume a prolate deformation. The intrinsic wave function is 

\begin{equation}
\Psi = (1-P_{12})d_{3/2, K=1/2}^{(1)} d_{3/2, K=-1/2}^{(2)}(1-P_{34}) f_{7/2, 1/2}^{(3)} f_{7/2, -1/2}^{(4)}
\end{equation}

We expand terms terms of states of ground total angular momentum. For a state of total angular momentum $I$ we have

\begin{equation}
\Psi = N\sum_{J_{P even} J_{N even}} (3/2\;\; 3/2\;\; 1/2\;\; -1/2\;\; \mid J_P\;\; 0) (7/2\;\; 7/2\;\; 7/2\;\; -7/2\;\; \mid J_N\;\; 0) (J_P\;\; J_N\;\; 0\;\; 0\;\; \mid J\;\; 0) \quad [J_P J_N]J
\end{equation}

Thus the wave function of the $2^+$ states is

\begin{equation}
\Psi (2^+) = \sum D(J_P J_N)[(d_{3/2}^2)J_P(f_{7/2}^2)J_N]^J
\end{equation}

\begin{equation}
D(J_P J_N) = N(3/2\;\; 3/2\;\; 1/2\;\; -1/2\;\; \mid J_P\;\; 0) (7/2\;\; 7/2\;\; 7/2\;\; -7/2\;\; \mid J_N\;\; 0) (J_P\;\; J_N\;\; 0\;\; 0\;\; \mid 2\;\; 0)
\end{equation}

Where $N$ is a normalization constant so that

\begin{equation}
\sum _{J_{P}J_{N}} |D(J_P J_N)|^2 = 1
\end{equation}

\subsection{$^{44}$Ar Prolate}

Here, the configuration is

\begin{equation}
\begin{gathered}
(1-P_{12}) d_{3/2,\; 1/2}^{(1)} \quad d_{3/2,\; -1/2}^{(2)} \\
(1-P_{34}) f_{7/2,\; 7/2}^{(3)} \quad f_{7/2,\; -7/2}^{(4)}
\end{gathered}
\end{equation}

where we find it more convenient to deal with two $f_{7/2}$ holes rather than six $f_{7/2}$ particles. Since the magnetic moment of a hole is the same as that of a particle, we can be somewhat cavalier. Hence,

\begin{equation}
D(J_P J_N) = N (3/2\;\; 3/2\;\; 1/2\;\; -1/2\;\; \mid J_P\;\; 0) (7/2\;\; 7/2\;\; 7/2\;\; -7/2\;\; \mid J_N\;\; 0) (J_P\;\; J_N\;\; 0\;\; 0\;\; \mid 2\;\; 0)
\end{equation}

\section{The Lawson prolate $g$ factors}

Here are the wave functions

\begin{center} Table I - Wave functions of $J = 2^+$ \\
states in the Lawson scheme \\
\begin{tabular}{| c | c | c |} \hline
& $^{40}$Ar (prolate) & $^{44}$Ar (prolate) \\ \hline
D(0,2) & 0.6485 & 0.7470 \\ \hline
D(2,0) & -0.5943 & 0.4890 \\ \hline
D(2,2) & -0.3466 & -0.3928 \\ \hline
D(2,4) & -0.3258 & 0.2085 \\ \hline
\end{tabular} \end{center}

The expression for the $g$ factor is 

\begin{equation}
g = \left(\frac{g_{\pi } + g_{\nu }}{2} \right) + \left(\frac{g_{\pi } - g_{\nu }}{2} \right) \sum _{J_P J_N} \frac{|D(J_P J_N)|^2 [J_P(J_P+1) - J_N(J_N+1)]}{J(J+1)}
\end{equation}

We find the following values

$a)$  Using bare $g$ factors - Prolate Lawson

\begin{equation*}
\begin{gathered}
g(^{40}\textrm{Ar}) = -0.3306 \\
g(^{44}\textrm{Ar}) = -0.3638
\end{gathered}
\end{equation*}

$b)$  Using empirical $g$ factors - Prolate Lawson

\begin{equation*}
\begin{gathered}
g(^{40}\textrm{Ar}) = -0.2102 \\
g(^{44}\textrm{Ar}) = -0.2479
\end{gathered}
\end{equation*}

The $^{40}$Ar result should be compared to what was obtained in a large shell model calculation \cite{stephan05, speidel08}.  It is somewhat more negative than the two measured values -0.02(4) and -0.1(1).  In a shell model calculation with the WBT interaction in the $d_{3/2}$ proton and $f_{7/2}$ neutron space the bare $g$ factor of the $2^+$ state is -0.441. This should be compared with the above Lawson value of -0.3306. The $g$ factors of $^{44}$Ar have not yet been measured so we here have predictions.

\section{Oblate Lawson}

It is easy to show

\begin{equation*}
\begin{gathered}
g(^{40}\textrm{Ar})_{Oblate} = g(^{44}\textrm{Ar})_{Prolate} \\
g(^{44}\textrm{Ar})_{Oblate} = g(^{40}\textrm{Ar})_{Prolate}
\end{gathered}
\end{equation*}

The reason for this is that it is equally valid to treat the proton wave function as that of 2 particles in $K = \pm 1/2$ orbits or alternatively as 2 proton holes in $K = \pm 3/2$ orbits. Note that the $^{40}$Ar and $^{44}$Ar results are not too different.

\section{An Unusual Shell Model Symmetry for $^{40}$Ar and $^{44}$Ar}

An interesting observation is that a shell modell calculation in the $d_{3/2\;\pi} \;\; f_{7/2\;\nu}$ space yields \uline{identical} results for the energy levels and $g$ factors of $^{40}$Ar and $^{44}$Ar. The reason is a bit subtle because in general a proton-neutron hole interaction is not the same as a proton-neutron interation. However, we are dealing with a system in which we have a half filled $d_{3/2}$ shell. We can regard the $^{40}$Ar calculation as one between two $d_{3/2}$ protons and two $f_{7/2}$ neutrons. We can regard the $^{44}$Ar calculation as one between two $d_{3/2}$ proton holes and two $f_{7/2}$ neutron holes. It is well known that the hole-hole interaction is to a constant, the same as the particle-particle interaction. This explains the reason that the energy levels and $g$ factors (and also B(E2)'s) are the same in this model space. Since the quadrupole moment of a hole is opposite to that of a particle, the results are also consistent with those in Sec. V in 'Oblate Lawson.' Any deviation of this symmetry would be evidence of configuration mixing.

\section{$^{46}$Ar}

In our model, $^{46}$Ar has the configuration $d_{3/2}^2 \quad f_{7/2}^8$. Since the neutron $f_{7/2}$ shell is closed we again get the result
\begin{equation*}
g(^{46}\textrm{Ar}) \approx g(^{39}\textrm{K})
\end{equation*}
Should we, in $^{39}$K, use the bare value or the measured value for $^{46}$Ar? If in $^{39}$K there is a significant contribution due to large intruder state conditions then the answer is not so clear. We note that $^{48}$Ca is a better clsoed shell than $^{40}$Ca. So it may be that for $^{46}$Ar we will get a $g$ factor closer to the bare value 0.084 rather than the 
effective value 0.28.

\section{Results for $J$ = 4, 6, and 8}

\begin{center}
\begin{tabular}{| c | c | c |}
\multicolumn{3}{c}{Table II - Wave functions of $J = 4^+$} \\
\multicolumn{3}{c}{states in the Lawson scheme}\\ \hline
& $^{40}$Ar & $^{44}$Ar \\ \hline
D(0 4) & 0.7350 & -0.5624 \\ \hline
D(2 2) & 0.5607 & 0.7723 \\ \hline
D(2 4) & 0.3749 & -0.2866 \\ \hline
D(2 6) & 0.0700 & 0.0696 \\ \hline
\multicolumn{3}{c}{ }\\ \hline
Bare $g$ & -0.4356 & -0.3519 \\ 
Empirical $g$ & -0.3829 & -0.2343 \\ \hline
\end{tabular}
\end{center}

See Section V for the oblate case. We now consider $J=6$ states

\begin{center}
\begin{tabular}{| c | c | c |}
\multicolumn{3}{c}{Table III - Wave functions of $J = 6^+$} \\
\multicolumn{3}{c}{states in the Lawson scheme}\\ \hline
& $^{40}$Ar & $^{44}$Ar \\ \hline
D(0 6) & -0.7282  & -0.3042 \\ \hline
D(2 4) & -0.5785 & 0.9401 \\ \hline
D(2 6) & -0.3674 & -0.1535 \\ \hline
\multicolumn{3}{c}{ }\\ \hline
Bare $g$ & -0.4701 & -0.3598 \\ 
Empirical $g$ & -0.3687 & -0.2433 \\ \hline
\end{tabular}
\end{center}

We now consider the unique case of $J=8$. The only configuration is $J_P = 2, J_N = 6$, hence D(2 6) =1.

\begin{equation}
g = \frac{(g_P+g_N)}{2} - \left( \frac{(g_P-g_N)}{2} \left(\frac{36}{72} \right) \right)
\end{equation}

\begin{equation*}
\begin{gathered}
g(^{40}\textrm{Ar}) = g(^{44}\textrm{Ar}) \\
\textrm{Bare value} = -0.3890\\
\textrm{Empirical value} = -0.2764\\
\end{gathered}
\end{equation*}

One can also obtain states of odd total angular momentum from the configurations that we have considered e.g. with $J_P =2$ and $J_N=4$ one can obtain $J=3^+$ and $J=5^+$.  From other configurations one can obtain $J=1^+$ and $J=7^+$.

\section{Clebsch-Gordon Coefficients}

\begin{equation*}
\begin{gathered}
(3/2\;\; 3/2\;\; 1/2\;\; -1/2\;\; \mid 0\;\; 0) = -0.5 \\
(3/2\;\; 3/2\;\; 1/2\;\; -1/2\;\; \mid 2\;\; 0) = 0.5 \\
(3/2\;\; 3/2\;\; 1/2\;\; -1/2\;\; \mid 6\;\; 0) = 0.307729 \\
(3/2\;\; 3/2\;\; 3/2\;\; -3/2\;\; \mid 0\;\; 0) = 0.5 \\
(3/2\;\; 3/2\;\; 3/2\;\; -3/2\;\; \mid 2\;\; 0) = 0.5 \\
(3/2\;\; 3/2\;\; 3/2\;\; -3/2\;\; \mid 6\;\; 0) = 0.061546 \\
(7/2\;\; 7/2\;\; 1/2\;\; -1/2\;\; \mid 0\;\; 0) = -0.353553 \\
(7/2\;\; 7/2\;\; 1/2\;\; -1/2\;\; \mid 2\;\; 0) = 0.385758 \\
(7/2\;\; 7/2\;\; 1/2\;\; -1/2\;\; \mid 4\;\; 0) = -0.362620 \\
(7/2\;\; 7/2\;\; 1/2\;\; -1/2\;\; \mid 6\;\; 0) = 0.307729 \\
(7/2\;\; 7/2\;\; 7/2\;\; -7/2\;\; \mid 0\;\; 0) = 0.353553 \\
(7/2\;\; 7/2\;\; 7/2\;\; -7/2\;\; \mid 2\;\; 0) = 0.540062 \\
(7/2\;\; 7/2\;\; 7/2\;\; -7/2\;\; \mid 4\;\; 0) = 0.282038 \\
(7/2\;\; 7/2\;\; 7/2\;\; -7/2\;\; \mid 6\;\; 0) = 0.061546 \\
(0\;\; 2\;\; 0\;\; 0\;\; \mid 2\;\; 0) = 1 \\
(2\;\; 0\;\; 0\;\; 0\;\; \mid 2\;\; 0) = 1 \\
(2\;\; 2\;\; 0\;\; 0\;\; \mid 2\;\; 0) = -0.534522 \\
(2\;\; 4\;\; 0\;\; 0\;\; \mid 2\;\; 0) = 0.534522 \\
(0\;\; 4\;\; 0\;\; 0\;\; \mid 4\;\; 0) = 1 \\
(2\;\; 2\;\; 0\;\; 0\;\; \mid 4\;\; 0) = 0.871371 \\
(2\;\; 4\;\; 0\;\; 0\;\; \mid 4\;\; 0) = -0.509646 \\
(2\;\; 6\;\; 0\;\; 0\;\; \mid 4\;\; 0) = 0.560968 \\
(2\;\; 4\;\; 0\;\; 0\;\; \mid 6\;\; 0) = 0.674200 \\
(0\;\; 6\;\; 0\;\; 0\;\; \mid 6\;\; 0) = 1 \\
(2\;\; 6\;\; 0\;\; 0\;\; \mid 6\;\; 0) = -0.504525 \\
\end{gathered}
\end{equation*}

\section{Closing Remarks}

We here present a simple model to qualitatively describe the $g$ factors of excited states in $^{40}$Ar, some of which has been measured, and in $^{44}$Ar, where there are no measurements thus far. Although ultimately large shell model calculations should be done (and in the cases of $^{38}$Ar and $^{40}$Ar they have) the method used here gives explicit wave functions so we can see the interplay of the contributors of neutrons and protons. We note in the expression for $g$ the factor $[J_P(J_P+1) -J_N(J_N+1)]$. This is largest when $J_P$ and $J_N$ are furthest apart. This amplifies the effect of $J_N = 4$ for say states of the total angular momentum $J=2$. Such a term is not present in `s-d models.' It is often surprising that in large shell model calculations that even when the percentage of the leading configurations (e. g. $d_{3/2\;\pi} \;\; f_{7/2\;\nu}$ in this case) is much less than one, the features of the simple shell model seem to hold. By comparing the small space calculations with large space ones, perhaps we can gain insight as to why it is so. Indeed with one of the interactions in \cite{stephan05} the leading configuration in only 70\% of the full wave function.

One of us (S. Y.) thanks the Aresty Program for support.

\bibliographystyle{unsrt}
\bibliography{Sources}

\begin{thebibliography}{1}

\bibitem{stephan05}
E.~A. Stephanova{,} N. Benczer{-}Koller{,} G. J. Kumbartzki{,} Y. Y. Sharon{,}
  L. Zamick{,} S. J. Q. Robinson{,} L. Bernstein{,} J. R. Cooper{,} D.
  Judson{,} M. J. Tayolr{,} M.~A. McMahan{,} and L.~Phair.
\newblock {\em Phys Rev}, 72:014309, 2005.

\bibitem{speidel06}
K.~{-}H. Speidel{,} S. Schielke{,} J. Leske{,} J. Gerber{,} P. Maier-Komor{,}
  S. J. Q. Robinson{,} Y. Y. Sharon{,}~L. Zamick.
\newblock {\em Phys Lett B}, 632:207, 2006.

\bibitem{speidel08}
K.~H. Speidel{,} S. Schielke{,} J. Leske{,} N. Pietralla{,} T. Ahn{,} A.
  Costin{,} O. Zell{,} J. Gerber{,} P. Maier{-}Komor{,} S. J. Q. Robinson{,} A.
  Escuderos{,} Y.~Y. Sharon{,} and L.~Zamick.
\newblock {\em Phys Rev}, 78:017304, 2008.

\bibitem{lawson61}
R.~D. Lawson.
\newblock {\em {Phys Rev}}, 124:1500, 1961.

\bibitem{lawson80}
R.~D. Lawson.
\newblock {\em Theory of the Nuclear Shell Model}.
\newblock Clarindar Press, Oxford, 1980.

\end{thebibliography}
\end{document}